# Effect of nano-carbon dispersions on signal in silicon-based sensor structure with photoelectrical transducer principle

Anton I. Manilov [a,b,*], Aleksey V. Kozinetz [a,b], Sergiy V. Litvinenko [a,b], Valeriy A. Skryshevsky [a,b], Mohammed Al Araimi [c,d], Alex Rozhin [c]

[a] Institute of High Technologies, Taras Shevchenko National University of Kyiv, Volodymyrska 64, 01033 Kyiv, Ukraine

[b] Corporation Science Park Taras Shevchenko University of Kyiv, Volodymyrska Str., 60, 01033 Kyiv, Ukraine

[c] Nanotechnology Research Group, Aston Institute of Photonic Technologies, School of Engineering & Applied Science, Aston University, Aston Triangle, B4 7ET Birmingham, United Kingdom

[d] Higher College of Technology, Al-Khuwair, PO Box 74, Postal Code 133, Sultanate of Oman

* Corresponding author. E-mail address: anmanilov@univ.kiev.ua (A.I. Manilov).

**Abstract**

We identified different nano-carbon species such as graphene nanoplatelets, graphite flakes and carbon nanotubes dispersed in N-methyl-2-pyrrolidone using a novel sensor structure based on a "deep" silicon barrier working as a photoelectrical transducer. Each nano-carbon particle has specific signature in both 2D photocurrent distribution and photocurrent dependences on bias changing surface band-bending. Additionally, all nano-carbon particles have characteristic features in the time-dependent evolution of photocurrent. The obtained results can be explained by the influence of nano-carbon molecules' local electric field on the recombination parameters of defect centers on the silicon surface.

## 1. Introduction

Nowadays, synthesis and research of carbon nano-materials (CNM) are the driving force for disruptive innovations in the field of nano-electronics, photonics, sensors, biomedical applications as well as in new composites with unprecedented mechanical, thermal and conductive properties [1-8]. Conventional nano-carbon species such as graphene, nanotubes, fullerenes and nanodiamonds have the physical and chemical properties determined by their geometric dimensions and the state of their surface [9-11]. Consequently, surface functionalization of carbon nanomaterials allows essentially changing the properties of the initial materials and makes possible interaction with different biological and chemical substances [5, 7, 12]. The latter is very promising for creating various biological and chemical sensors, as well as for detecting the nanoparticles themselves [13-17]. Despite such sensors are less sensitive than methods of analytical chemistry, they benefit from fast and convenient measurements, the ability to transfer information to computer servers and mobile platforms almost instantly.

There are a lot of chemical sensors based on electrical, optical or photoluminescent conversion principles. One of the widely known possible approaches that can be used for detecting analyte with complex molecular structure is LAPS (Light Addressable Potentiometric Sensor) [18]. In previous works we consider photoelectrical sensor with a bit different structure but another transducer mechanism: the "deep" silicon barrier is illuminated by strongly absorbed light [19-20]. It has been successfully implemented to probe carbon nanotubes with adsorbed surfactant in an aqueous medium [21]. The technical realization of this method is easier comparing to LAPS. Here, the application of the structure with the same operational principle for detection of different nano-carbon species in organic solvent is analyzed.

## 2. Experimental procedure

Three types of carbon nanostructures are explored: carbon nanotubes (CNT), graphite flakes (GF) and graphene nanoplatelets (GNP). All carbon samples were dispersed in N-methyl-2-pyrrolidone (NMP).

Commercially available high-pressure carbon monoxide method (HiPCO) single-walled carbon nanotubes were purchased from Unidym (Lot # P0261). The SWNTs' dispersion was prepared by placing 1 mg of HiPCO SWNTs in 10 mL NMP. The mixture was then sonicated, using a commercial ultrasonic processor for one hour at 200W and 20 kHz. Then, the resultant mixture was immediately ultracentrifuged for 2 h 30 minutes under 17 °C, using MLS 50 rotor for 45000 revolutions per minute (RPM) to remove impurities and residual bundles. After centrifugation, the top 70% of the resultant mixture was used as initial SWNTs' dispersion.

Graphite flakes were produced by Sigma-Aldrich® (PubChem Substance ID 24859898). Dispersion of GF (0.5 g) in NMP (10 mL) was prepared by ultrasonication at 21 kHz and 250 W. The resulting mixture was then centrifuged at 17 °C for one hour at 30000 RPM. Graphene nanoplatelets were obtained from Acros Organics[TM] (Fisher Scientific UK Ltd). Dispersion of GNP (150 mg) in NMP (10 mL) was made by ultrasonication and ultracentrifugation at the same conditions as for the GF mixture.

The nanomaterials concentrations were controlled by measuring the optical absorption of dispersions at 660 nm. Further analytical calculation using the Beer-Lambert law gave concentrations for CNT - 0.004 mg/ml, while for GF and GNP - 0.06 and 0.07 mg/ml respectively. At such concentrations, the dispersions of all the samples showed stability without visible aggregation within 3-4 months.

Size characterization of the nanoparticles in NMP solvent was realized by dynamic light scattering and zeta potential measurements using Zetasizer Nano S (Malvern Instruments). The samples were irradiated with a helium-neon laser with λ = 633 nm at 25 °C.

The silicon sensor structure with photoelectrical transducer principle is suggested for the detection of dispersed carbon nanomaterials in this study. The method is based on measurement of laser beam induced photoelectric signal formed on "deep" barrier of silicon wafer with the presence of a liquid analyte on its surface (fig. 1).

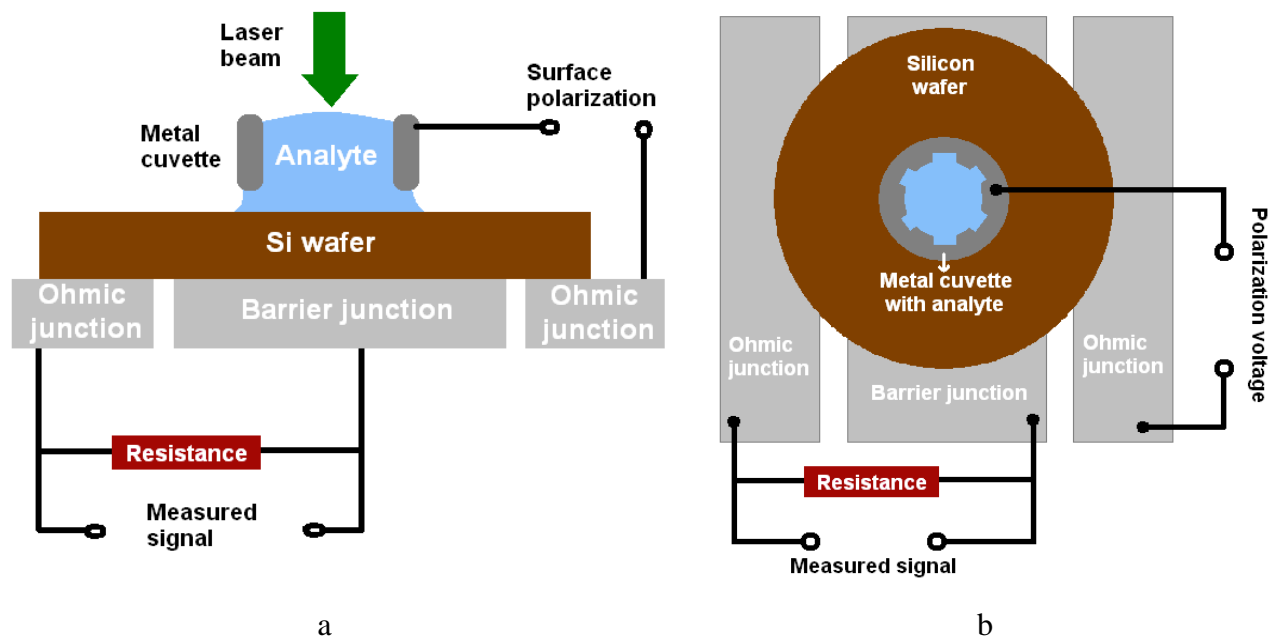

Fig. 1. Measurement scheme: side view (a) and front view (b)

The wafers are double-side polished n-Si (100) with specific resistivity 30-50 Ω·cm. The samples are electrically connected to three bottom contacts. Barrier junction is provided on the central contact. The side contacts are prepared for Ohmic junctions. The amplitude modulated green laser (532 nm) is used for illumination. Potential barrier for separation of photo-generated charge carriers is created by Schottky metal-semiconductor contact. The liquid analyte is held on the surface of the wafer using a metal cuvette (volume is 0.1 mL). Also, the metal cuvette plays the role of an electrode for surface polarization, as indicated in fig. 1. A small gap between the cuvette and silicon surface is made to make sure that the analyte does not propagate beyond the working cell. This can take place due to the surface tension of water and NMP.

There are two measurement modes: surface scanning and surface polarization. The surface scanning is using a focused laser beam and results in surface distribution of photoelectric signal at constant surface polarization. The areas inside and outside of the cuvette are explored by this way. The surface polarization is realized by measurement of photoelectric signal with different external voltages applied to the analyte using with a fixed position of the illuminated spot (the spot size is 0.1 mm). The laser illumination area in this mode is at the center of the cuvette.

## 3. Results

The characteristic distributions of the carbon particles sizes, received from Zetasizer, are given in Fig. 2. As can be seen from the obtained curves, GNP particles' sizes are mostly within the range 40-60 nm. Maximum for GF particles distribution is between 80-110 nm. Two maximums are observed for CNT: centered nearly 100 nm and 800 nm. The measured values of zeta potential are -17 mV, -12 mV and -19 mV for GNP, GF and CNT respectively. The negative values of zeta potential are typical for carbon nanoparticles in organic solvent [22]. It should be mentioned that these values suggest relative stability of nanoparticles in the solvent at temperature 25 °C. Some instability and coagulation tendency are possible in the case of abrupt change of external conditions.

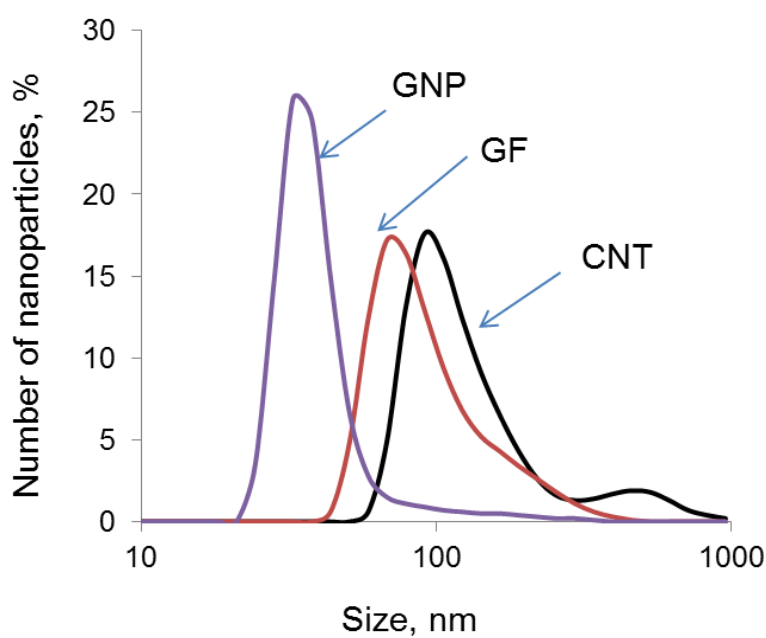

Fig. 2. The size distributions of the nano-carbon species in NMP.

Figure 3 shows the photocurrent distribution of the laser scanning over the sensor surface where X and Y axis represent coordinates. The polarization voltage was not applied to the surface during the scanning measurements. The photocurrent increases after addition of the analyte in the cuvette (fig.3, a,b). The similar regularity is observed for pure NMP and all dispersions of carbon nanomaterials. After 30-60 min from placing of the analyte on the Si surface the photocurrent inside the cuvette starts behaving differently depending on the type of the sample. The dispersions of carbon nanotubes and graphene nanoplatelets, as well as pure NMP, are characterized with relatively high photocurrent during the measurement (fig.3, d). At the same time the samples of graphite flakes show some decreasing of the signal that is especially noticeable comparing to area outside the cuvette (fig.3, c). This effect is repeated for all GF samples.

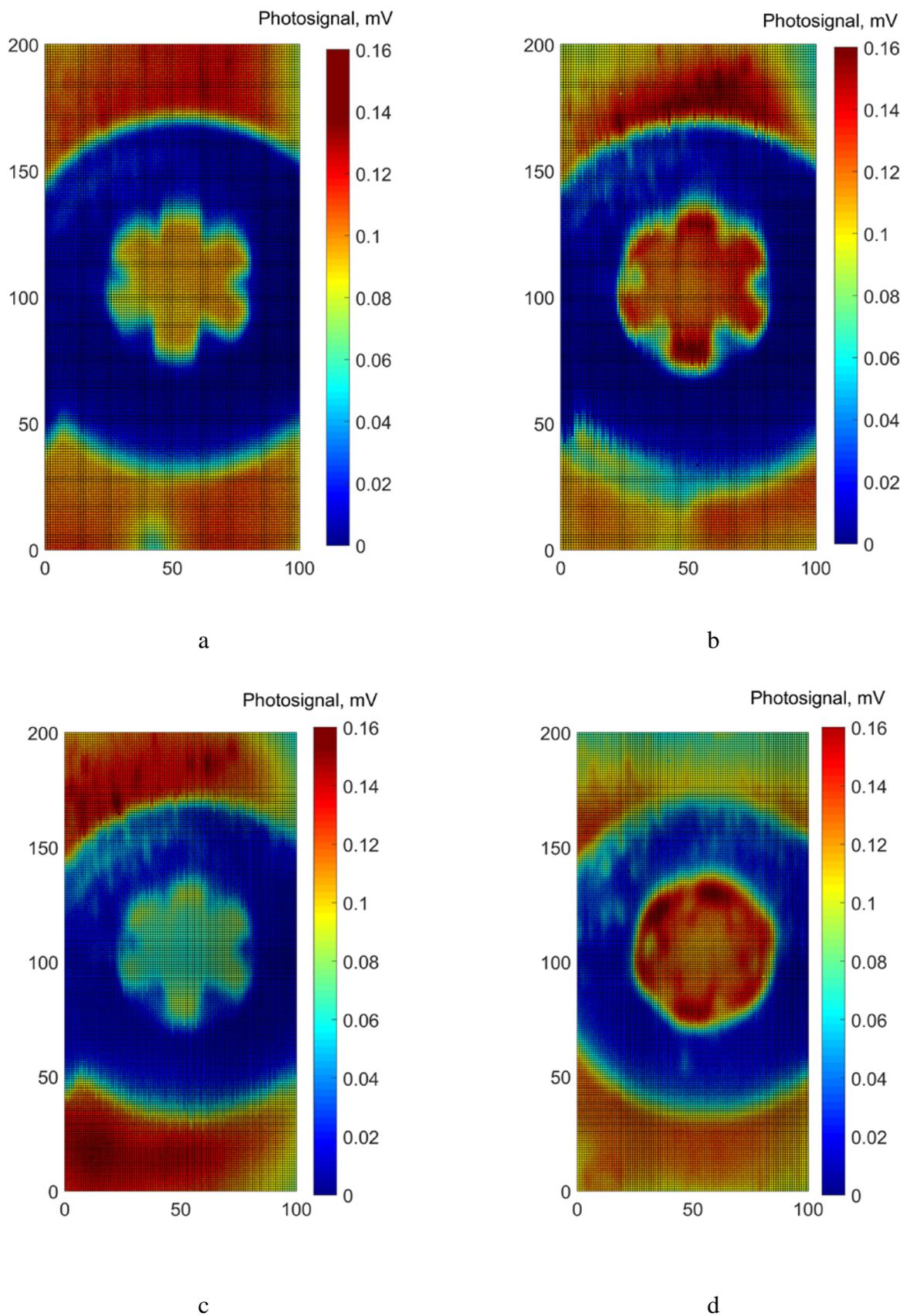

Fig.3. Photosignal surface distribution on n-Si wafer with the cuvette filled with the analyte:

(a) empty cuvette;

(b) typical photocurrent distribution for all samples based on NMP solvent after 10 min on the surface;

(c) typical photocurrent distribution for GF dispersions after 90 min on the surface;

(d) typical photocurrent distribution for CNT and GNP dispersions, as well as pure NMP, after 90 min on the surface.

The scanned area is about 1x2 cm, XY axis values represent coordinates of the scanned points.

The surface polarization measurements are represented by the series of bell-shaped curves for different analytes (fig.4). Each curve is photocurrent dependence on voltage applied between the cuvette electrode and bottom contact of the Si wafer. The series are formed by periodic measurements during 90 min from addition of the analyte in the cuvette.

Apparently, the samples based on NMP solvent have wider minimums and lower slope of the right branches of the bell-shaped curves in comparison with the graphs of water analyte. Moreover, the left branches on the plots of the samples based on NMP have no minimums observed on the curves corresponding for water.

The measured photocurrent dependencies on surface polarization voltage for the nano-carbon dispersions differ from the plot obtained for pure NMP solvent (fig. 4, e). Moreover, shape of the curves and their modification with time are distinctive for each type of the carbon nanostructure. All GF samples are characterized with shift of left branch of the "bell" to the lower voltages during the aging of the analyte on the semiconductor surface (fig. 4, d). This is accompanied with widening of the curve's minimum. Besides, it results in decrease of photocurrent at zero potential as observed during surface scanning (fig.3, c). The effect is not observed for CNT and GNP samples (fig.4, b,c, fig.3, d).

Also, the left branches of the curves usually alter during the measurement (fig.4, b,d), excluding CNT samples that are characterized with relatively stable left branch (fig. 4 c).

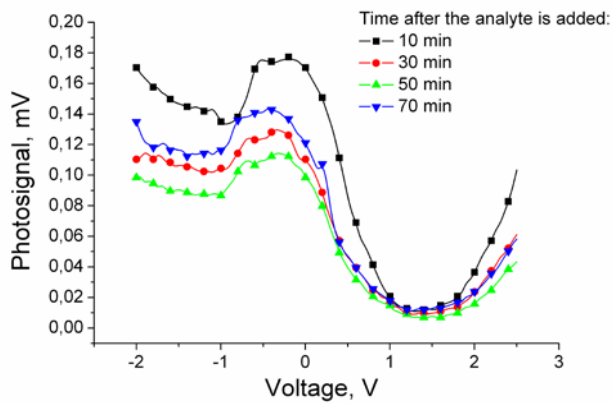

a

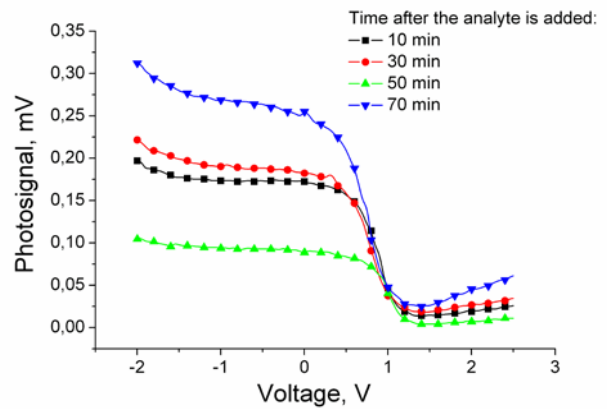

b

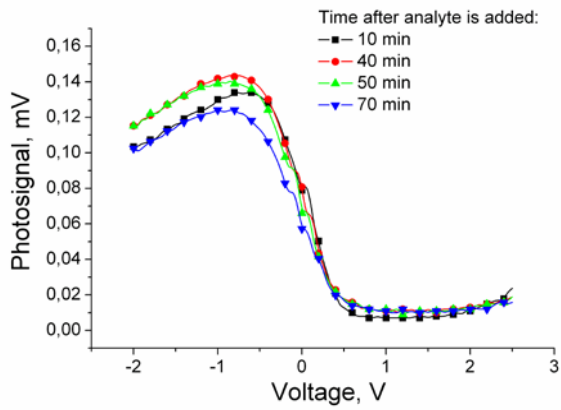

c

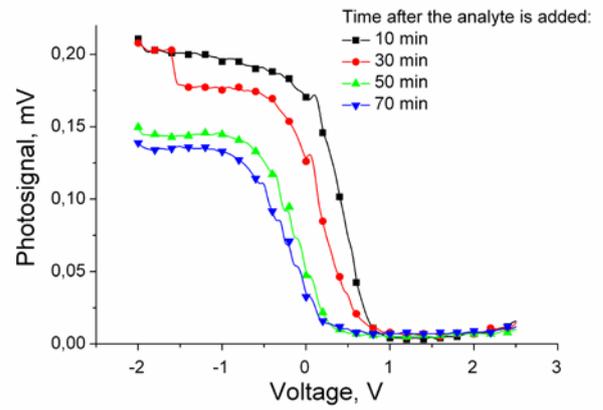

d

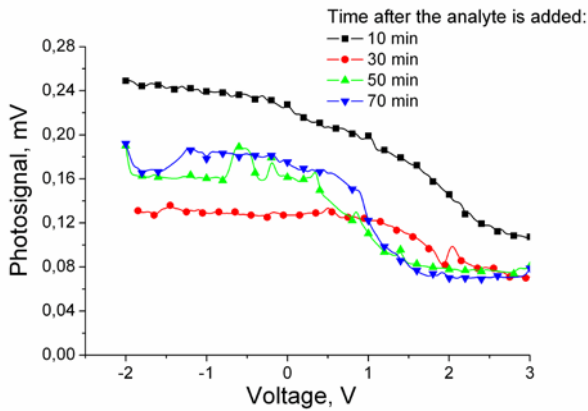

e

Fig.4. Photocurrent dependencies on voltage applied between the cuvette with analyte and bottom contact of Si wafer. The next analytes are displayed:

(a) distilled water;

(b) GNP solution in NMP;

(c) CNT solution in NMP;

(d) GF solution in NMP;

(e) pure NMP.

## 4. Discussion

The electrical scheme of our measurements (fig.1) allows to realize the "deep" silicon p-n junction configuration (or more generally any silicon barrier structure) with illumination by the laser beam [19]. Let us consider some aspects of the signal formation in the scheme. The measured signal is proportional to photocurrent of the barrier structure. The photocurrent is collected due to built-in electric field of space charge region. It is shown in [20] that deep silicon junction can be regarded as effective photoelectrical transducer when light is utilized with strong absorption coefficient. The wavelength of the laser illumination belongs to strong light absorption in silicon $\alpha(\lambda) \sim 10^5\text{-}10^3$ cm$^{-1}$. In our case the thickness of $n$-$Si$ wafer $d$ is nearly 300 μm, holes diffusion length $l_p$ is nearly 150μm, so when the surface of base region is illuminated all principal conditions of "deep" junction configuration are fulfilled. It means that the region, where non-equilibrium holes are generated ($1/\alpha(\lambda) \sim$ μm), and region, where these carriers are collected by electric field, are separated in space. The photocurrent $I$ in the deep structure significantly depends on surface recombination velocity $S$, it can be given as:

$$I(S) \cong \frac{1 + \dfrac{S}{\alpha(\lambda)D_p}}{S\dfrac{l_p}{D_p}sh\left(\dfrac{d}{l_p}\right) + ch\left(\dfrac{d}{l_p}\right)} \quad , \tag{1}$$

where $D_p$ is hole diffusion coefficient.

It means that change of $S$ by one or two order of its magnitude leads to the change of photocurrent by several dozen times. This fact is illustrated in Fig.5.

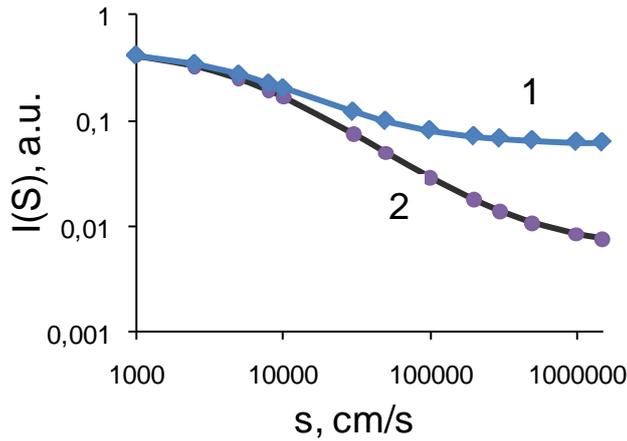

Fig.5 The calculated dependences (1) of the photocurrent of deep silicon barrier structure on surface recombination velocity $S$ for $\alpha(\lambda) \sim 10^3$ cm$^{-1}$, $\lambda = 630$nm (1), and $\alpha(\lambda) \sim 10^4$ cm$^{-1}$, $\lambda = 532$ nm (2).

The photocurrent depends slightly on the recombination velocity for shallow junction (up to several μm), because the region of built-in electric field is in the vicinity of illuminated surface.

The parameter $S$ corresponds to energy values of recombination levels, its concentration, recombination cross sections for electrons and holes and surface band bending $\varphi_s$ in the frame of Stevenson- Keyes theory and it more complicated modification [23].

In principle, this parameter can be different for the analytes with different molecular dipole structures. So, the first mode, measured for the analyte, is photocurrent 2D distribution (that is determined by effective surface recombination velocity for fixed initial surface band bending).

On the other hand, applying voltage $V$ with different polarity (via surface polarization circuit) allows changing the surface band bending $\varphi_s$ in the region of silicon wafer that is covered with analyte. The principal traits of the Stevenson- Keyes theory claim that if the concentration of non-equilibriums electrons and holes are close to each other, the $S$ gets its maximal value, if default of electrons or holes takes place, $S$ is strongly diminished. That is why the dependences $S(\varphi_s)$ typically have bell-shaped form. As it was mentioned earlier, laser beam can be positioned in any point of the surface that contacts with analyte. Surface polarization in the scheme (fig.1) allows, in some sense, imaging of the curve $S(\varphi_s)$ into the curve $I(V)$ for fixed point at the surface. Thus, the easily measured curve *I(V)* can be the second characteristic mode of the analyte. In some sense, the described set-up realizes sensor structure with photoelectrical (recombination-based) transducer.

Let us consider experimental results presented in figures 3-4. The initial increase of photocurrent (fig. 3 a,b) is most likely explained by change of surface band bending in the region of silicon wafer contacting with the analyte (inside cuvette). It can be connected with decrease of $S$ and/or passivation of recombination centers at the silicon surface. Comparing with free silicon surface, the photocurrent for the silicon area covered with the nano-carbon dispersion increases by two times.

As can be seen in fig. 4, the minima of curves *I(V)* (maximums of $S(\varphi_s)$) correspond to positive voltage applied to the surface of *n*-silicon for all samples. Since the base region has conductivity of n-type, the depletion region appears near illuminated surface. The minimum of the dependences *I(V)* is in range of +1 V – +1.5 V for all cases. Consequently, the recombination cross sections for electrons and holes should remain practically unchanged.

Let us consider the curve received for initial moment of observation and the curves received for subsequent moments (indication of time is given at the right corners of fig. 4, a-e). Let us consider the curves deviations from initial one. For each of the cases corresponding to fig. 4, the amplitudes and character of the deviations are clearly different. As can be seen, these effects are observed namely for the left branches of the experimental curves. When the Si surface contacts with water (fig. 4 a), the shape of the curve *I(V)* varies insignificantly with time. Thus the recombination occurs basically via the level with stable recombination parameters at the interface analyte-*n*-Si. The peculiarities of *I(V)* near -1 V at the left branch of the curve show that recombination process may be more complex. Rather the system of energy levels at the silicon surface (local or distributed with energy) is possible and several channels of the recombination can be realized.

The carbon samples based on NMP dispersion have noticeably wider minima and lower slopes of the right branches of the bell-shaped curves comparing to the case of neat DI water. The widening of the curves can be explained by change of the energy position of the recombination level. It is well known that the greater is the difference between the recombination energy level and the middle of the silicon band gap, the wider is the bell of $S(\varphi_s)$ and *I(V)*.

The change of photocurrent in time varies for the different nano-carbon samples. The GNP dispersion in fig. 4 b demonstrates increase of *I(V)* (left branch) with time. So, the decrease of recombination levels concentration at the surface is in progress. The curves for CNT (fig4, c) demonstrate certain similarity with the neat distilled water, because no evolution of the dependences *I(V)* is observed in time. The *I(V)* curves for graphite flakes (fig.4,d) become wider with time that can be explained by the gradual shift of recombination level energy from the middle of silicon band gap. This effect is supplemented with decrease of the photocurrent at zero bias (fig. 3,c). Moreover, the photocurrent in the saturation region (i.e. when applied voltage is less than -1 V) clearly indicates increasing of the recombination value with time. The growth of $S$ is explained by increase of the recombination centers' concentration on the surface. So, two types of the changes are observed simultaneously.

As can be seen, the shapes of the curves $I(V)$ and their evolution in time, as well as the aging regularities of the photocurrent surface distributions, are very specific for each of the analytes that have been investigated. The chemical and electrostatic aspects should be taken into account to clarify the situation.

It is known that degradation of NMP solvent occurs during ultrasonication by way of ring opening and supplementary chemical reactions [24]. The resulting organic substances are adhered onto the surface of nano-carbon species. As can be seen from Fig 2, graphite flakes, carbon nanotubes and graphene nanoplatelets are of different size and volume. Also, they have different structure of the surface and corresponding covering particularities. Together with different level of catalytic features of nanoparticles these effects can influence the chemical activity of the functionalized nano-carbon and the solvent itself. It can result in different time dependences of the analyte-silicon interaction.

The chemical reaction of the analyte at the silicon surface influences pre-surface band-bending and the recombination properties of the silicon surface. On the other hand, when silicon/analyte interface is illuminated, non-equilibrium carriers are generated in the $n$-silicon wafer. It can result in the minority carrier moving to the interface (due to the electric field of space charge region near the interface). Moreover, these carriers can be involved in the surface chemical reactions. That is why silicon surface passivation can be facilitated or hindered with time.

The electrostatic interaction of the analyte molecules with the silicon surface is specific for each of four investigated cases. The electrical charge formed on the surface of nano-carbon is result of adsorption of the organic compounds from the solution and/or the dissociation of the surface compounds. Thus, each type of carbon nanoparticle carries some electrical charge distributed on its surface. In that way, specific configuration of dipole structures is formed for each of nanoparticle type in the solution. This configuration strongly depends on the relative spatial orientation of charged fragments. The local electrical field of the analyte molecules is probably the main factor that determines change of the initial parameters of the recombination centers at the surface of $n$-Si (that allows the detection in principle). It can be either local recombination center or system of recombination centers at the surface that define photocurrent behavior in our experiments. The energy levels can be non-uniformly distributed at the silicon surface because the dipoles of the surrounding analyte molecules cause the thermal fluctuations of polarization.

So, the silicon sensor structure with photoelectrical transducer principle can be applied for differentiation of the nano-carbon species in NMP.

## 5. Conclusions

The possibility of detecting graphite flakes (GF), carbon nanotubes (CNT) and graphene nanoplatelets (GNP) in N-methyl-2-pyrrolidone (NMP) solvent is demonstrated. The approach is based on the analysis of the photocurrent surface distribution and the photocurrent dependences on voltage (changing surface band bending) of the "deep" silicon barrier structure.

Addition of the nano-carbon dispersion on the sensor results in the increase of the photocurrent in two times comparing with free silicon surface. The dependences of photocurrent on voltage are distinctive for each analyte: the case of GNP corresponds to narrow curve; the CNT and GF are characterized with wider curves. The signal for GNP samples in saturation region is in 1.5-2 times greater than signal for CNT and GF dispersion.

The additional information is also received from the time-dependent measurements of the photocurrent. The GNP dispersion is characterized with irregular in time photocurrent at negative and zero polarizing bias. The photocurrent for the GF samples decreases at zero bias. This is accompanied with widening of the I(V) curve's minimum at aging of the analyte on the surface. The relatively stable behavior is characteristic for CNT samples.

These effects can be explained by the influence of the local molecular electrical field on the recombination parameters of the defect levels on the silicon surface (i.e. recombination levels energies, their concentration, recombination cross sections and pre-surface band bending). It can be either local level in the silicon band gap or system of levels continuously distributed with the energy. On the basis of the experimental results, it can be maintained that local electric field of each analyte changes the parameters of the recombination centers in its own specific way.

These allow consider the proposed methods of the nano-carbon detection for further investigation and application.

**Acknowledgements**

This work was supported by EU Horizon 2020 Research and Innovation Staff Exchange Programme (RISE) under Marie Sklodowska-Curie Action (project 690945 "Carther"). M.A. acknowledges the support from the Ministry of Higher Education, Sultanate of Oman.

## References

[1] C. Gao, G. Chen, Conducting polymer/carbon particle thermoelectric composites: Emerging green energy materials, Compos. Sci. Technol. 124 (2016) 52-70.

[2] D. Liu, W. Zhao S. Liu, Q. Cen, Q. Xue, Comparative tribological and corrosion resistance properties of epoxy composite coatings reinforced with functionalized fullerene C60 and graphene, Surf. Coat. Technol. 286 (2016) 354–364.

[3] M. Cen-Puc, A.I. Oliva-Avilés, F. Avilés, Thermoresistive mechanisms of carbon nanotube/polymer composites, Physica E 95 (2018) 41-50.

[4] K. Tsukagoshi, N. Yoneya, S. Uryu, Y. Aoyagi, A. Kanda, Y. Ootuka, B.W. Alphenaar, Carbon nanotube devices for nanoelectronics, Physica B 323 (2002) 107-114.

[5] C.S. Park, H. Yoon, O. S. Kwon, Graphene-based nanoelectronic biosensors, J. Ind. Eng. Chem. 38 (2016) 13-22.

[6] L. Sun, X. Wang, Y. Wang, Q. Zhang, Roles of carbon nanotubes in novel energy storage devices, Carbon 122 (2017) 462-474.

[7] A. Nag, A. Mitra, S.Ch. Mukhopadhyay, Graphene and its sensor-based applications: A review, Sens. Actuators, A 270 (2018) 177-194.

[8] R. Wang, L. Xie, S. Hameed, C. Wang, Y. Ying, Mechanisms and applications of carbon nanotubes in terahertz devices: A review, Carbon 132 (2018) 42-58.

[9] K.D. Sattler, Carbon nanomaterials sourcebook. Volume 1. Graphene, fullerenes, nanotubes, and nanodiamonds, CRC Press, Taylor & Francis Group, 2016.

[10] N. A. Abdel Ghany, S. A. Elsherif, H. T. Handal, Revolution of Graphene for different applications: State-of-the-art, Surfaces and Interfaces 9 (2017) 93-106.

[11] J. Coro, M. Suárez, L.S.R. Silva, K.I.B. Eguiluz, G.R. Salazar-Banda, Fullerene applications in fuel cells: A review, Int. J. Hydrogen Energy 41 (2016) 17944-17959.

[12] F.V. Ferreira, L. Cividanes, F.S. Brito, B.R.C. de Menezes, W. Franceschi, E.A. Nunes Simonetti, G.P. Thim, Functionalizing Graphene and Carbon Nanotubes, SpringerBriefs in Applied Sciences and Technology, 2016.

[13] M.E. Foo, S.C.B. Gopinath, Feasibility of graphene in biomedical applications, Biomed. Pharmacother. 94 (2017) 354-361.

[14] M. I. Sajid, U. Jamshaid, T. Jamshaid, Carbon nanotubes from synthesis to in vivo biomedical applications, Int. J. Pharm. 501 (2016) 278-299.

[15] K. Bates, K. Kostarelos, Carbon nanotubes as vectors for gene therapy: Past achievements, present challenges and future goals, Adv. Drug Delivery Rev. 65 (2013) 2023-2033.

[16] M. Ema, M. Gamo, K. Honda, A review of toxicity studies on graphene-based nanomaterials in laboratory animals, Regul. Toxicol. Pharm. 85 (2017) 7-24.

[17] C. Bussy, L. Methven, K. Kostarelos, Hemotoxicity of carbon nanotubes, Adv. Drug Delivery Rev. 65 (2013) 2127-2134.

[18] J. Wang, L. Du, S. Krause, C. Wu, P. Wang, Surface modification and construction of LAPS towards biosensing applications, Sens. Actuators, B 265 (2018) 161-173.

[19] S.V. Litvinenko, A.V. Kozinetz, V.A. Skryshevsky. Concept of photovoltaic transducer on a base of modified p-n junction, Sens. Actuators, A 224 (2015) 30-34.

[20] A.V. Kozinetz, S.V. Litvinenko, V.A. Skryshevsky. Physical properties of silicon sensor structures with photoelectric transformation on the basis pacs of "deep" $p$–$n$-junction, Ukr. J. Phys. 62 (2017) 318-325.

[21] A.I. Manilov, A.V. Kozinetz, I.V. Gavrilchenko, Y.S. Milovanov, T.M. Mukhamedzhanov, S.A. Alekseev, M. Al Araimi, S.V. Litvinenko, A. Rozhin, V.A. Skryshevsky, Photoelectric signal conversion in deep p-n junction for detection of carbon nanotubes with adsorbed SDBS in aqueous solution, J. Nano- Electron. Phys. 9 (2017) 04020.

[22] I. Swyzer, R. Kaegi, L. Sigg, B. Nowack, Colloidal stability of suspended aqnd agglomerate structures of settled carbon nanotubes in different aqueous matrices, Water Res. 47 (2013) 3910-3920.

[23] D.T. Stevenson, R.J. Keyes, Measurements of the recombination velocity at germanium surfaces, Physica 20 (1954) 1041-1046.

[24] H.Ch. Yau, M.K. Bayazit, J.H.G. Steinke, M.S.P. Shaffer, Sonochemical degradation of N-methylpyrrolidone and its influence on single walled carbon nanotube dispersion, Chem. Commun. 51 (2015) 16621.